\documentclass[useAMS,usenatbib]{mn2e}
\usepackage{graphicx}
\title[V371 Per]{V371 Per - A Thick-Disk, Short-Period F/1O Cepheid}
\author[P. Wils et al.]{P. Wils$^{1}$, A.A. Henden$^{2}$, S. Kleidis$^{3,4}$, E.G. Schmidt$^{5}$, D.L. Welch$^{6}$\\
$^{1}$Vereniging voor Sterrenkunde, Belgium, email: patrickwils@yahoo.com \\
$^{2}$American Association of Variable Star Observers, 49 Bay State Road, Cambridge, MA 02138, USA, email: arne@aavso.org \\
$^{3}$Zagori Observatory, Epirus, Greece, email: steliosklidis@gmail.com \\
$^{4}$Helliniki Astronomiki Enosi, Athens, Greece \\
$^{5}$Department of Physics and Astronomy, University of Nebraska, Lincoln, Nebraska 68588, USA, email: eschmidt1@unl.edu \\
$^{6}$Dept. of Physics \& Astronomy, McMaster University, Hamilton, Ontario, L8S 4M1 Canada, email: welch@physics.mcmaster.ca \\
}

\begin{document}

\date{Accepted 2009 October 27.  Received 2009 October 27; in original form 2009 September 11}

\pagerange{\pageref{firstpage}--\pageref{lastpage}} \pubyear{2009}

\maketitle

\label{firstpage}

\begin{abstract}
V371~Per was found to be a double-mode Cepheid with a fundamental mode period of 1.738 days, 
the shortest among Galactic beat Cepheids, and an unusually high period ratio of 0.731, 
while the other Galactic beat Cepheids have period ratios between 0.697 and 0.713.
The latter suggests that the star has a metallicity [Fe/H] between -1 and -0.7.  
The derived distance from the Galactic Plane places it in the Thick Disk or the Halo,
while all other Galactic beat Cepheids belong to the Thin Disk.
There are indications from historical data that both the fundamental and first overtone periods have lengthened.
\end{abstract}


\section{Introduction}

The variability of V371~Per was discovered by \citet{weber}.  
From photographic studies, \citet{satyvaldiev} and \citet{meinunger} did not find any periodicity, and the star was therefore
assumed to be an irregular variable.
\citet{behlen} found a period of 1.2697 days from CCD data and suspected it to be a beat Cepheid because of its changing light curve.
This interpretation agrees with the spectral type G0 given by \citet{bond}.
As a consequence, the star was observed for a number of years at the Behlen Observatory at the University of Nebraska,
the United States Naval Observatory Flagstaff Station (NOFS), 
the Sonoita Research Observatory (SRO; Sonoita, Arizona) and
the Zagori Observatory (Athens, Greece).  The resulting data will be presented in this paper.
In addition to the new data, the data sets from \citet{satyvaldiev} and \citet{behlen} are reanalysed
(the latter with revised values for the comparison stars), 
and data from the Northern Sky Variability Survey \citep[NSVS,][]{nsvs} are analysed as well.

\section{Observations}

Table~\ref{log} contains a summary of the new observations.  
All data will be made available at the Centre de Donn\'ees astronomiques de Strasbourg (CDS).
A comparison star sequence in the Johnson-Cousins system, derived from observations at SRO, is given in Table~\ref{comps}.
Observations on each photometric night included following an extinction star from low to high airmass, along with $BVR_CI_C$ exposures of Landolt standard fields.
Further details on the procedure are outlined in \citet{wvir}.
A similar procedure was used to obtain the data from the NOFS.  
The comparison star sequence was chosen to have a large colour range in the vicinity of the variable.  
GSC~3854-1439 is a close double of about $1\arcsec$ separation of which both components need to
be included in any aperture.

\begin{figure*}
\includegraphics[width=\textwidth]{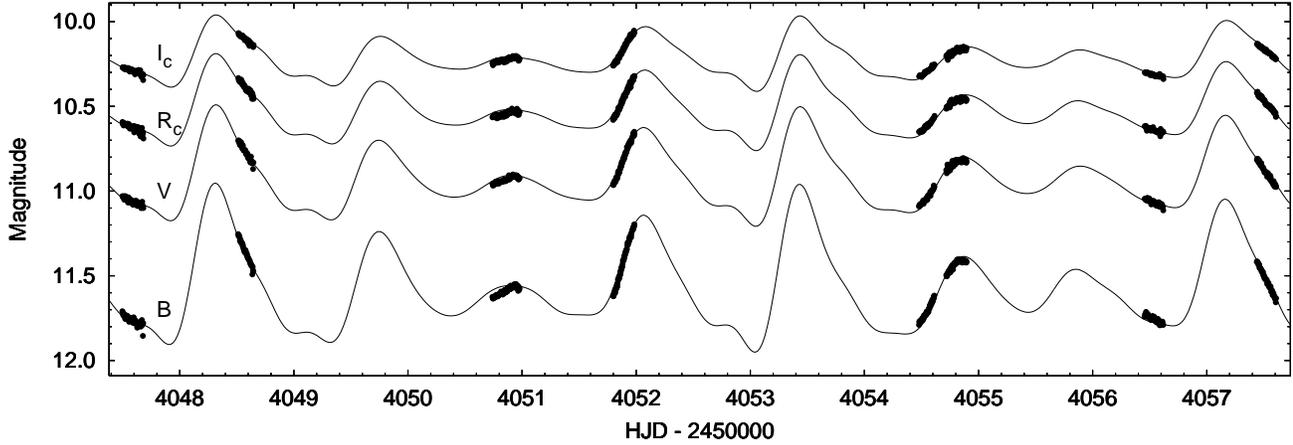}
 \caption{Part of the light curves and model curves in $BVR_cI_c$ for V371~Per.}
 \label{plot}
\end{figure*}

The method described in \citet{behlenobs} was used to reduce the data from the Behlen Observatory.
Data from the Zagori Observatory were reduced with the photometry package AIP4Win \citep{aip} 
and were transformed to the standard system with transformation coefficients obtained by measuring several of the comparison stars from Table~\ref{comps}.
While the colour terms for $V$ and $I_c$ were found to be almost negligible, 
those for $B$ and $R_c$ involved corrections of up to 0.05 magnitude.
Fig.~\ref{plot} contains a section of the data superposed on a model plot as derived in the following section.

\begin{table*}
 \caption{Observation log for V371~Per.}
 \label{log}
 \begin{tabular}{lllcrr}
  \hline
Observatory  & Instruments     & Filters & JD - 2400000 & \multicolumn{1}{c}{Nights} & \multicolumn{1}{c}{Points} \\ 
  \hline
Behlen & 76-cm Cassegrain + TI4849          & $VR_c$   & 47872-53967 & 51 &  157 \\
NOFS   & 1.0-m Ritchey-Chr\'etien + Tektronix & $BVR_cI_c$ & 50033-51528 & 32 &   47 \\ 
SRO    & 35-cm C-14 + SBIG STL 1001E        & $BVR_cI_c$ & 53629-54054 & 23 & 1341 \\ 
Zagori & 30-cm LX200 + SBIG ST-7XMEI        & $BVR_cI_c$ & 53726-54135 & 38 & 1331 \\ 
  \hline
 \end{tabular}
\end{table*}

\begin{table*}
 \caption{Comparison star data from the Sonoita Research Observatory.}
 \label{comps}
 \begin{tabular}{lllrrrrrrrr}
  \hline
GSC ID    & \multicolumn{2}{c}{RA (J2000) Dec} & \multicolumn{1}{c}{$V$} & \multicolumn{1}{c}{$\sigma_V$}
          & \multicolumn{1}{c}{$B-V$} & \multicolumn{1}{c}{$\sigma_{B-V}$}
          & \multicolumn{1}{c}{$V-R_c$} & \multicolumn{1}{c}{$\sigma_{V-R}$}
          & \multicolumn{1}{c}{$R_c-I_c$} & \multicolumn{1}{c}{$\sigma_{R-I}$} \\ 
  \hline
2854-0440 & 02:54:43.43 & +42:27:07.8 & 11.700 & 0.016 & 0.634 & 0.026 & 0.373 & 0.013 & 0.289 & 0.013 \\
2854-0492 & 02:55:05.14 & +42:38:07.7 & 12.262 & 0.013 & 0.741 & 0.023 & 0.427 & 0.018 & 0.340 & 0.016 \\
2854-0965 & 02:55:05.88 & +42:39:10.9 & 11.704 & 0.010 & 0.471 & 0.017 & 0.275 & 0.020 & 0.202 & 0.022 \\
2854-1439 & 02:55:14.45 & +42:33:24.2 & 13.145 & 0.003 & 0.596 & 0.007 & 0.354 & 0.005 & 0.377 & 0.007 \\
2854-1056 & 02:55:15.05 & +42:32:35.6 & 12.793 & 0.004 & 0.952 & 0.006 & 0.529 & 0.006 & 0.493 & 0.008 \\
2854-0637 & 02:55:23.95 & +42:45:05.3 & 12.339 & 0.013 & 0.709 & 0.031 & 0.406 & 0.021 & 0.339 & 0.022 \\
2854-1058 & 02:55:37.11 & +42:30:36.8 & 11.570 & 0.020 & 1.453 & 0.021 & 0.799 & 0.011 & 0.713 & 0.021 \\
2854-0696 & 02:55:37.35 & +42:44:21.6 & 11.922 & 0.012 & 0.673 & 0.032 & 0.387 & 0.021 & 0.301 & 0.021 \\
2854-0713 & 02:55:45.27 & +42:31:16.5 &  9.609 & 0.036 & 0.185 & 0.041 & 0.095 & 0.035 & 0.127 & 0.035 \\
2854-0768 & 02:56:03.32 & +42:43:01.7 & 10.728 & 0.013 & 0.511 & 0.029 & 0.307 & 0.023 & 0.224 & 0.011 \\
2854-1013 & 02:56:09.96 & +42:39:52.4 & 11.658 & 0.007 & 0.479 & 0.020 & 0.289 & 0.012 & 0.210 & 0.012 \\
  \hline
 \end{tabular}
\end{table*}

\section{Period analysis}

A Fourier analysis of the data was done using Period04 \citep{period04}.
For all available data sets, two independent frequencies were found: 
a first one, hereafter referred to as $f_1$, near 0.787~c/d or 1.270~d, corresponding to the period found by \citet{behlen}, 
and a second one, further referred to as $f_0$, near 0.576~c/d or 1.737~d.
The ratio $f_0/f_1=0.7312$ also proves that the suggestion of a beat Cepheid was correct 
and that $f_0$ and $f_1$ should be interpreted respectively as the frequency of the fundamental and the first overtone mode. 

\begin{figure}
\includegraphics[width=\columnwidth]{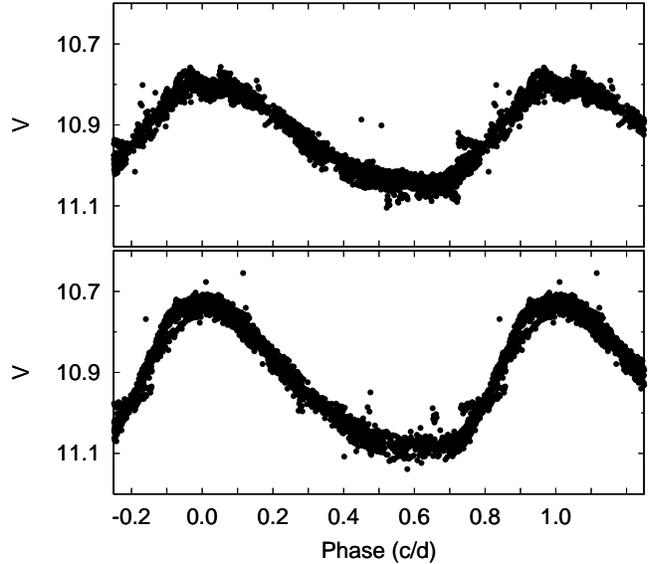}
 \caption{Phase plot of $V$ data from SRO and Zagori Observatory.  
The top panel shows the data prewhitened for the first overtone mode and all combination frequencies as given in Table~\ref{freq}. 
The bottom panel shows the data prewhitened for the fundamental mode and all combination frequencies.}
 \label{phase}
\end{figure}

The two most extensive data sets also reveal a number of combination modes of these two frequencies,
demonstrating that the variations observed are from a single star and not, for example, from two
separate pulsators within the same resolution element on the sky,
as those would not show these combination frequencies.
The details of the frequencies found are given in Table~\ref{freq}.  
For each passband and detected frequency the semi-amplitude, phase and signal-to-noise ratio are given.
The values of the frequencies themselves were derived from the $V$ data, 
and then used to calculate the amplitudes and phases for the other passbands.
The small decrease in phase when going to redder colours for $f_0$ and $f_1$ indicate that these are radial modes.
Phase plots of the fundamental mode variation and the first overtone variation in $V$ 
(prewhitened for the other mode and combination modes) are given in Fig.~\ref{phase}.

\begin{table*}
 \caption{Frequencies for V371~Per and their semi-amplitudes (in millimag) and phases (in degrees) derived from the SRO and Zagori data sets.  
Uncertainties calculated from Monte Carlo simulations with Period04 \citep{period04}, are given between parenthesis in units of the last decimal.}
 \label{freq}
 \begin{tabular}{llr@{}lrrrrrrrrrrr}
  \hline
\multicolumn{2}{c}{Frequency (c/d)} & \multicolumn{2}{c}{$A_B$} & \multicolumn{1}{c}{$\phi_B$} & $S/N_B$ & 
                                      \multicolumn{1}{c}{$A_V$} & \multicolumn{1}{c}{$\phi_V$} & $S/N_V$ &
                                      \multicolumn{1}{c}{$A_R$} & \multicolumn{1}{c}{$\phi_R$} & $S/N_R$ & 
                                      \multicolumn{1}{c}{$A_I$} & \multicolumn{1}{c}{$\phi_I$} & $S/N_I$ \\ 
  \hline
$f_1$ & 0.787329(3) & 244.5 & (10) & 314(1) & 51.1 & 173.5(6) & 311(1) & 68.1 & 136.2(6) & 309(1) & 45.5 & 105.4(5) & 305(1) & 56.0 \\
$f_0$ & 0.575678(4) & 175.2 & (10) & 109(1) & 60.7 & 124.8(7) & 106(1) & 63.5 & 99.2(6) & 102(1) & 50.7 & 77.2(4) & 97(1) & 44.8 \\
$f_0+f_1$ & 1.363007 & 85.8 & (11) & 189(1) & 23.9 & 60.0(7) & 189(1) & 29.0 & 48.4(7) & 186(1) & 18.9 & 37.3(4) & 186(1) & 22.9 \\
$2f_1$ & 1.574658 & 53.3 & (11) & 44(2) & 14.6 & 38.6(6) & 40(1) & 17.0 & 30.0(6) & 41(2) & 12.5 & 23.2(5) & 41(2) & 12.8 \\
$f_1-f_0$ & 0.211651 & 41.7 & (9) & 152(2) & 9.8 & 27.5(6) & 154(2) & 11.5 & 24.1(6) & 148(2) & 7.5 & 19.4(5) & 154(2) & 8.5 \\
$2f_0$ & 1.151355 & 25.2 & (8) & 11(3) & 4.8 & 18.3(7) & 356(2) & 5.8 & 13.6(7) & 360(3) & 3.7 & 11.4(4) & 360(3) & 4.5 \\
$2f_0+f_1$ & 1.938684 & 23.1 & (9) & 115(3) & 4.2 & 17.9(7) & 110(2) & 7.2 & 14.9(6) & 117(3) & 4.3 & 12.8(4) & 122(3) & 5.9 \\
$f_0+2f_1$ & 2.150336 & 34.0 & (10) & 303(2) & 7.2 & 24.9(7) & 306(2) & 10.0 & 19.0(6) & 306(2) & 5.9 & 15.3(5) & 307(2) & 7.4 \\
$3f_1$ & 2.361987 & 14.1 & (9) & 173(4) & 4.4 & 11.7(6) & 160(4) & 6.0 & 10.6(6) & 167(4) & 4.9 & 9.1(5) & 162(4) & 5.6 \\
$3f_0$ & 1.727033 & 16.3 & (8) & 298(4) & 3.8 & 7.4(7) & 302(5) & 4.2 & 9.9(5) & 304(4) & 3.7 & 7.6(5) & 312(4) & 4.0 \\
  \hline
 \end{tabular}
\end{table*}

The amplitude of the first overtone mode is larger than that of the fundamental mode, 
which is not common among Galactic beat Cepheids.
Only for AX~Vel (with a fundamental period of 3.67~d) and V458~Sct (4.84~d), 
has the first overtone mode with a larger amplitude than the fundamental pulsation mode.
There does not seem to be a relation of amplitude ratio with period.

\begin{figure}
\includegraphics[width=\columnwidth]{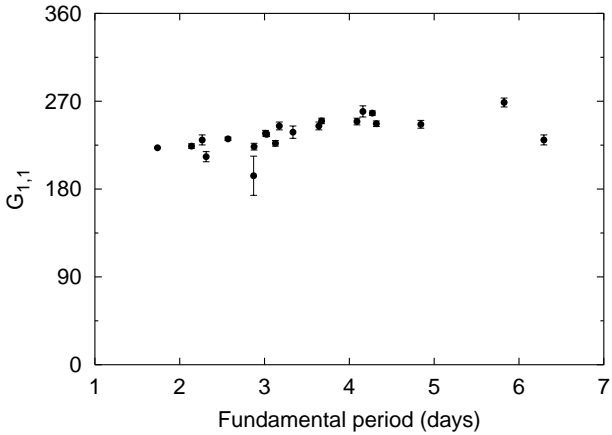}
 \caption{Values (in degrees) of the generalized phase difference $G_{1,1}$ for the cross coupling term $f_0+f_1$ for Galactic double-mode Cepheids.  
Besides V371~Per, the values are taken from \citet{poretti} and \citet{ibvs} or calculated from ASAS-3 data \citep{asas} when not available otherwise.}
 \label{g11}
\end{figure}

The value of the generalized phase difference $G_{1,1}=3.88\pm0.01$ (expressed in radians) 
for the cross coupling term $f_0+f_1$ \citep{poretti} 
is in agreement with those of the other Galactic beat Cepheids (see Fig.~\ref{g11}).
Note that the definition of $G_{1,1}$ in \citet{poretti} uses a cosine expansion for the Fourier series, while Period04,
and therefore also Table~\ref{freq}, uses a sine expansion, 
so that $\pi/2$ should be added to the result calculated from the data in Table~\ref{freq}.  
Only the star with the longest period, V367~Sct, does not follow the linear trend in $G_{1,1}$ with period.

\begin{figure}
\includegraphics[width=\columnwidth]{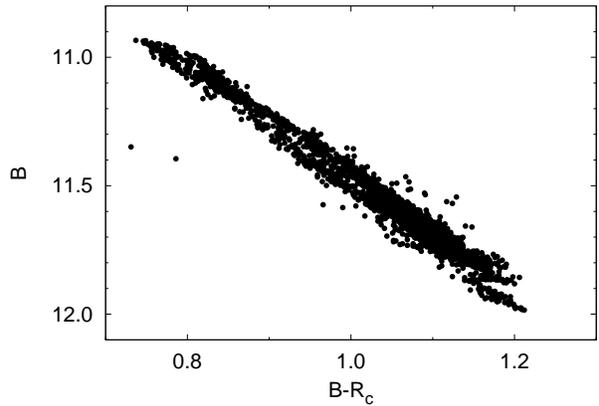}
 \caption{$B$ data from SRO and Zagori Observatory plotted against colour $B-R_c$.}
 \label{colour}
\end{figure}

There seems to be no indication for excess emission at any particular brightness or phase, as a plot of $B-R_c$ against $B$ shows in Fig.~\ref{colour}.

\section{Period evolution}

The values and semi-amplitudes of the fundamental and first overtone frequencies and the frequency ratio calculated from the other data sets
are given in Table~\ref{historical}.  For reference, the corresponding values already given in Table~\ref{freq} are repeated.
The periods derived from the NSVS data are less reliable because of the short time span of the data (less than a year).

\begin{figure}
\includegraphics[width=\columnwidth]{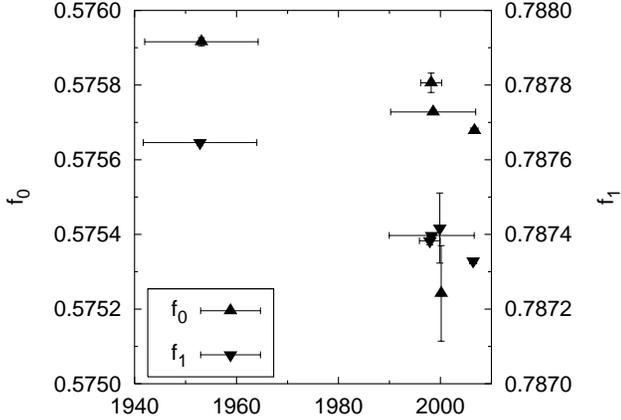}
 \caption{Frequency evolution of the fundamental and first overtone mode of V371~Per over the years from the data sets discussed in this paper.  
The horizontal bars indicate the years in which the observations took place.}
 \label{period}
\end{figure}

The derived frequency values are plotted in Fig.~\ref{period}.
There are indications that both the fundamental mode and first overtone frequency have decreased, 
while the frequency ratio remained constant at 0.7312.
This result highly depends on the photographic data set of \citet{satyvaldiev}, and should therefore be treated with caution.
However there is also a significant difference between the frequencies derived from the Behlen data set and those from the
newer SRO and Zagori data sets, in line with this conclusion.

\begin{table*}
 \caption{Fundamental and first overtone mode calculated for all data sets. 
Semi-amplitudes are given in millimag.}
 \label{historical}
 \begin{tabular}{lclllll}
  \hline
Data set & Filter & $f_0$ & \multicolumn{1}{c}{$A_0$} & $f_1$ & \multicolumn{1}{c}{$A_1$} & \multicolumn{1}{c}{$f_0/f_1$} \\ 
  \hline
Satyvaldiev & - & 0.575916(11) & 211(35) & 0.787646(7) & 295(37) & 0.73119(2) \\
Behlen & $V$ & 0.575728(5) & 116(8) & 0.787397(4) & 171(6) & 0.73118(1) \\
Behlen & $R_c$ &   & ~90(7) &   & 132(6) \\
NOFS & $V$ & 0.575806(26) & ~90(5) & 0.787383(11) & 148(5) & 0.73129(2) \\
NOFS & $B$ &   & 129(7) &   & 220(8) \\
NOFS & $R_c$ &   & ~66(6) &   & 114(6) \\
NOFS & $I_c$ &   & ~90(13) &   & 162(14) \\
NSVS & - & 0.575242(128) & 103(5) & 0.787417(94) & 140(5) & 0.73054(18) \\
SRO+Zagori & $V$ & 0.575678(4) & 125(7) & 0.787329(3) & 174(6) & 0.73118(1) \\
SRO+Zagori & $B$ &  & 175(1) &  & 245(1)  \\
SRO+Zagori & $R_c$ &  & ~99(1) &  & 136(1)  \\
SRO+Zagori & $I_c$ &  & ~77(1) &  & 105(1)  \\
  \hline
 \end{tabular}
\end{table*}

There is no indication that the amplitude has changed over the years. 
Note that the \citet{satyvaldiev} data are photographic and the amplitude should therefore be compared to the $B$ band amplitudes.
The NSVS data are unfiltered CCD data, so that amplitudes derived from them are to be compared with $R$ band amplitudes.
The lesser number of data points in the NOFS data sets give less reliable amplitudes, and their uncertainties are likely underestimated.

\section{Discussion}

\begin{figure}
\includegraphics[width=\columnwidth]{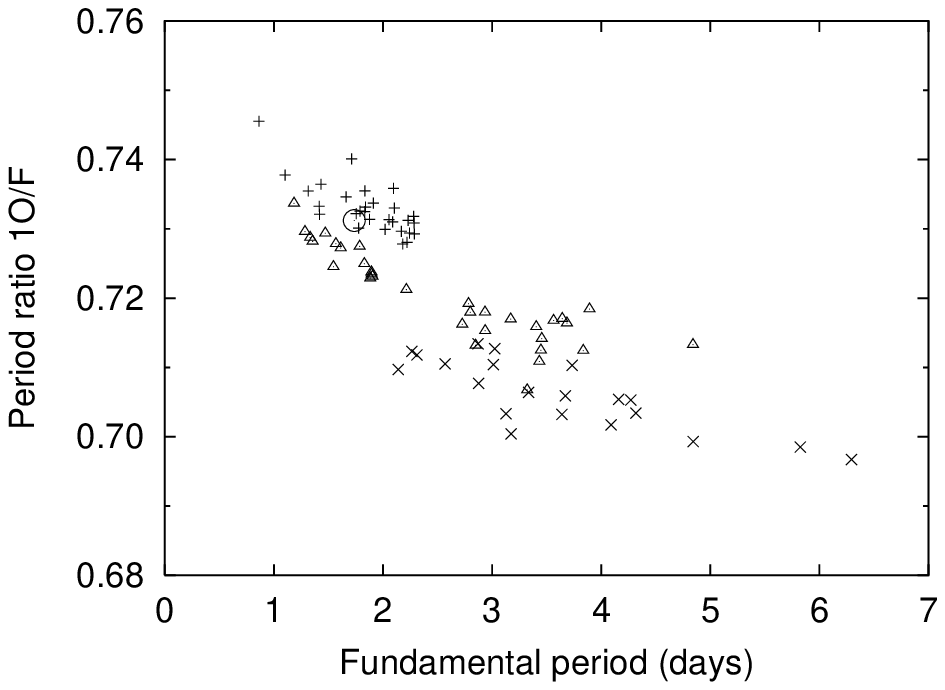}
 \caption{Period ratio $P_1/P_0$ plotted against fundamental period $P_0$ for Galactic Cepheids (shown as crosses; data from \citet{welch}, \citet{ibvs} and \citet{antipin};
the value for DZ~CMa has been calculated from ASAS-3 data \citep{asas}),
LMC Cepheids (triangles; data from \citet{macho-lmc} and \citet{ogle-lmc}) and SMC Cepheids (plusses, data from \citet{eros-smc}, \citet{macho-smc} and \citet{ogle-smc}).
V371~Per is depicted with a large circle.}
 \label{ratio}
\end{figure}

The fundamental period of 1.737~days of V371~Per is short and outside the range of the periods of the other known Galactic beat Cepheids.  
Until now, those with the shortest fundamental periods known were TU~Cas with a period of 2.139~days \citep{tucas} 
and DZ~CMa with a period of 2.311~days \citep[determined from ASAS-3 data, ][]{asas}.  
The beat Cepheid with the largest known period is V367~Sct \citep[6.293~days, ][]{v367sct}.
As can be seen in Fig.~\ref{ratio}, the period and period ratio of V371~Per are more reminiscent of the Small Magellanic Cloud double-mode Cepheids.  
From theoretical calculations, \citet{buchler} have shown a relation between fundamental period, period ratio and metallicity.
From the graphs of \citet{buchler} and \citet{buchler2} one can determine $Z$ to be approximately between 0.002 and 0.004 for V371~Per, 
depending on the particular mixture of elements chosen for $Z$,
equivalent to $[Fe/H]$ between -1 and -0.7.  
From the empirical formula (3) given by \citet{spectroscopy}, 
also relating fundamental period and period ratio to metallicity for Galactic beat Cepheids,
a value $[Fe/H]=-0.92$ can be derived by extrapolation.  
Both values are in good agreement and indeed indicate a low metallicity, uncommon for a Galactic Cepheid,
as the values cited by \citet{spectroscopy} range from $[Fe/H]=-0.20\pm0.14$ for VX~Pup to $[Fe/H]=0.09\pm0.18$ for V458~Sct.
This is also low compared to Galactic Cepheids pulsating in the fundamental mode only, as these have $-0.5<[Fe/H]<0.4$ \citep{luck,lemasle,pedicelli}.
Even taking into account a metallicity gradient with Galactocentric distance, does not allow for this discrepancy.
An independent estimate for the metallicity of V371~Per can be deduced by following the procedure detailed by \citet{caputo1} for Cepheids pulsating in the fundamental mode.
This gives a significantly larger value of $Z\approx0.01$ or $[Fe/H]\approx -0.3$.
Note however that the relations computed by \citet{caputo1} are based on Cepheid masses larger than $5 M_{\sun}$, and periods longer than $\approx$ 3 days,
so that strictly speaking they do not hold for V371~Per.  
For example the quadratic relations in $log P$ for the blue and red edge of the instability strip intersect at $Z\approx0.011$ for the fundamental period of V371~Per,
and the colour-colour relation results in a negative extinction value for $Z=0.02$.

Furthermore, the Galactic latitude of V371~Per of -14.7 degrees is rather high, 
higher than for the other double-mode Cepheids in the Galaxy.
Apart from TU~Cas (at latitude +11.4), all the others have latitudes between -8 and +5.5.
$\eta$ Aql, a bright and therefore relatively nearby Cepheid, has a Galactic latitude of -13.7 degrees.

Although the metallicity of V371~Per is low and the height above the Galactic Plane is large, 
it is likely not a low mass Population~II variable.
The graphs of \citet{buchler} suggest a mass for V371~Per just above $3 M_{\sun}$, while BL~Her stars have masses smaller than $1 M_{\sun}$.
The light curve of V371~Per does not resemble those of BL~Her stars (Population~II Cepheids) with similar periods, as they have very steep rising branches.
In addition no double-mode Population~II Cepheids are known. 
Furthermore theoretical models from \citet{blher} suggest that there are no stable first overtone pulsations in BL~Her stars
with the fairly cool temperature of V371~Per.
With the extinction $E(B-V) = 0.11$ towards V371~Per estimated from \citet{extinction} (see Table~\ref{mag}),
its unreddened $B-V$ colour can be calculated as 0.54.  
The effective surface temperature should therefore be 6000K or lower \citep[depending on metallicity and surface gravity; ][]{temperature}.  
All of the models in \citet{blher} have the red edge of stable first overtone modes at temperatures above 6000K.

The proper motion of V371~Per is $4.1\pm1.1$~mas/yr from the UCAC2 catalogue \citep{ucac2}, much lower than some BL~Her stars
of comparable apparent magnitude and period (in mas/yr: BL~Her $13.2\pm1.4$, SW~Tau $10.8\pm1.4$, RT~TrA $14.1\pm2.5$, DU~Ara $12.3\pm5.1$, UY~Eri $25.7\pm1.8$).
Of course, center-of-mass radial velocities, combined with the proper motion and distance, will only
determine the kinematic group to which V371~Per belongs.

With the extinction data from Table~\ref{mag},
the empirical $BVR_cI_c$ period-luminosity (PL) relations given by \citet{pl} can be used to calculate the distance to V371~Per.
The average distance modulus derived from these four passbands is $m-M=12.51\pm0.07$,
corresponding to a distance of $3.2\pm0.1$~kpc, and a Galactocentric distance of 10.6~kpc.
Assuming the Sun is very near the Galactic Plane \citep{reed}, 
a height of 0.8~kpc above the Galactic Plane can then be calculated for V371~Per, 
much higher than all other Galactic Cepheids, which all lie within 300~pc from the Galactic Plane.  
This would place V371~Per above the Thin Disk, so that it is located in the Galactic Thick Disk or Halo,  
while all of the previously known Galactic beat Cepheids lie within the Galactic Thin Disk.
Using the PL relations for the reddening-free Wesenheit magnitudes given in \citet{pl}, 
a somewhat smaller distance modulus $m-M=12.34\pm0.03$ is obtained, 
bringing V371~Per about 50~pc closer to the Galactic Plane.
The 2MASS $JHK_s$ magnitudes \citep{2mass} give a similar distance modulus $m-M=12.33\pm0.03$
(care should be taken here because the exact pulsation phase of the 2MASS measurements is difficult to determine).

Using the calculated absolute magnitude from Table~\ref{mag}, an independent estimate can be made for the mass of V371~Per, 
using the relations for fundamental mode Cepheids established by \citet{caputo5}.
In principle these are only valid for stars with $Z=0.02$ and $P>3$ days.
Depending on the passband a pulsation mass between $2.3$ and $3.1 M_{\sun}$ is obtained.
With the luminosity taken to be the canonical luminosity $L_{can}$ \citep{caputo5}, the deduced values for the evolutionary mass 
range between $3.3$ and $3.8 M_{\sun}$ from the Mass-Period-Luminosity relation 
and between $3.6$ and $3.7 M_{\sun}$ from the Mass-Colour-Luminosity relation.
Although these estimates show a large range they do confirm V371~Per to be an intermediate-mass pulsator.

\begin{table}
 \caption{Mean apparent magnitude, extinction estimated from \citet{extinction} 
and absolute magnitude calculated from \citet{pl} for V371~Per in the $BVR_cI_c$ (our data)
and 2MASS $JHK_s$ passbands \citep{2mass}.}
 \label{mag}
 \begin{tabular}{cccc}
  \hline
Passband & Apparent &  Extinction & Absolute \\ 
         & mag. $m_\lambda$   & $A_\lambda$ & mag. $M_\lambda$ \\ 
  \hline
$B$ & 11.58 & 0.46 & -1.48 \\
$V$ & 10.93 & 0.35 & -1.92 \\
$R_c$ & 10.53 & 0.29 & -2.22 \\
$I_c$ & 10.22 & 0.21 & -2.44 \\
$J$ & 9.63 & 0.10 & -2.83 \\
$H$ & 9.35 & 0.06 & -3.01 \\
$K_s$ & 9.29 & 0.04 & -3.09 \\
  \hline
 \end{tabular}
\end{table}

\section{Conclusion}

Extensive CCD photometry of V371~Per over a number of years has clearly shown it to be a Galactic beat Cepheid, 
with the shortest period known so far.
The high value of the frequency ratio (0.731) suggests that its metallicity is much lower than that of the other
Galactic beat Cepheids.  
Its distance derived from empirical PL relations places it in the Galactic Thick Disk or Halo, 0.8~kpc above the Galactic Plane.
V371~Per is therefore a remarkable object and 
a spectroscopic study is warranted to confirm its low metallicity.
It is very likely that V371~Per will help to refine the theoretical models for Cepheid pulsation.
Further photometric observations may also confirm the period increase suspected for both modes.
This will become evident within the next five to ten years.

\section*{Acknowledgments}
Klaus H\"au\ss{}ler is acknowledged for providing a copy of the Satyvaldiev and Meinunger papers.
This study used data from the Northern Sky Variability Survey 
created jointly by the Los Alamos National Laboratory and the University of Michigan,
and funded by the US Department of Energy, 
the National Aeronautics and Space Administration (NASA) and the National Science Foundation (NSF).
Use of the SIMBAD and VizieR 
databases operated at the Centre de Donn\'ees astronomiques de Strasbourg \citep{vizier}
and the SAO/NASA Astrophysics Data System is gratefully acknowledged.
DLW acknowledges support from the Natural Sciences and Engineering Research Council of Canada (NSERC).
The authors are grateful to the reviewer, Giuseppe Bono, for constructive remarks to improve the paper.


\label{lastpage}

\end{document}